# Phosphorene Junctions as a Platform for Spin-Selective Quantum Dots in Next-Generation Devices

Maryam Mahdavifar [1,2], Farhad Khoeini[1]*, François M. Peeters [3,4]

[1]Department of Physics, Univesity of Zanjan, P.O. Box 45195-313, Zanjan, Iran

[2]Faculty of Physics, University of Tabriz, Tabriz, Iran

[3]Centre for Quantum Metamaterials, HSE University, Moscow 101000, Russia

[4]Departamento de Física, Universidade Federal do Ceará, Fortaleza, Ceará, 60455-760, Brazil

**ABSTRACT**

The impact of vacancies on spin-resolved electronic properties of quantum dots (QDs) in phosphorene-based junctions, are investigated numerically. Regardless of the crystal orientation, a phosphorene nanoribbon (PNR) containing a monovacancy is found to exhibit a topological quasi-flat band that emerges within the band gap. The electronic properties of QDs, including spatial confinement and energy level distribution, can be strongly tuned by controlling the topological structure of the QDs and by applying electric fields. Additionally, these QDs exhibit remarkable spin-selective properties under a ferromagnetic exchange field, enabling the manipulation of QD features. This opens up the potential for novel applications such as quantum computing, magnetic sensing, spin-based light emission.

**Keywords**: Phosphorene Junctions; Spatial Confinement; Topological Structure; Quantum Dots; Band Gapp; Electric Fields.

## INTRODUCTION

In the quest for novel layered materials with atomic thickness, single- or few-layer black phosphorus (BP), known as phosphorene, has been successfully fabricated by exfoliating bulk crystals [1, 2]. Phosphorene has a puckered honeycomb structure and shows a direct band gap that is highly thickness-dependent, increasing from ~0.3 eV for bulk to ~2 eV for monolayers [3, 4]. Studies on the electronic structure and optical properties of monolayer phosphorene have shown that there is in-plane anisotropy for two distinct zigzag and armchair directions [5]. The field effect transistor based on phosphorene shows a moderate on/off ratio of ~$10^4$, and carrier mobility as high as ~1000 $cm^2V^{-1}s^{-1}$ at room temperature [6-8].

Recent experiments show that it is possible to prepare high-quality phosphorene nanoribbons (PNRs) with various widths [9, 10]. The edges of the PNRs play a critical role in their electronic properties. PNRs generally show well-separated valence and conduction bands, resulting from the non-edge atoms. However, a defining characteristic of zigzag PNRs (ZPNRs) is the existence of a quasi-flat band within the bandgap, attributed to the zigzag edge [11]. Armchair PNRs (APNRs) have a semiconducting character, whereas non-passivated ZPNRs show metallicity stemming from the edge states.

Structural defects such as dislocations and vacancies are unavoidable in synthesized materials, which have a substantial influence on their properties [12-14]. In this work, we evaluate the effect of monovacancies on the electronic structure of PNRs. Our numerical results show that a monovacancy in a PNR, regardless of the crystal orientation, creates a quasi-flat band that is isolated from the bulk states and appears in the middle of the gap. Thus, the introduction of topological monovacancies in the structure of an APNR changes its electronic character from semiconducting to metallic. This suggests that a variety of semiconductor/semiconductor, semiconductor/metal, and metal/metal junctions can be obtained by introducing topological vacancies in an APNR.

One can expect that a quantum dot device can be created by merging two semiconductor/metal junctions [15, 16]. The study of the spin-resolved transport properties of phosphorene-based QD junctions and their response to a transverse electric field are topics of our letter. The electron spin in QDs has been suggested to provide a controllable realization of scalable qubits [17, 18]. Our calculations are based on the tight-binding (TB) model and the Green's function method. The results show that the spatial confinement and the localized energy level distribution are strongly influenced by the topological structure of the QD, which can also be tuned by a transverse electric field.

The manuscript is organized as follows. The next section presents the model and methodology. The third section presents the results and discussion. Finally, the conclusion summarizes the main findings.

## MODEL AND METHOD

### Atomic Structure and Tight-Binding Model

The atomic structure of phosphorene is shown in figure 1(a). The phosphorus atoms are evenly distributed between two half layers, forming a puckered honeycomb structure. A TB model has been proposed for phosphorene by introducing hopping parameters ($t_k$) over five neighboring sites [19] displayed in a sketchy way in figure 1(a). The specific values of these parameters are as follows: $t_1 = -1.220$ eV, $t_2 = 3.665$ eV, $t_3 = -0.205$ eV, $t_4 = -0.105$ eV, and $t_5 = -0.055$ eV. In addition, we use a model Hamiltonian for the spin-orbit coupling in phosphorene, which has been developed in Ref. [20].

### Calculation details

To study the transport properties of the junctions, we divided the device into three regions: a middle region (scattering region) connected to two semi-infinite segments. The effective Hamiltonian arises from the coupling of the middle region with the side connections. The contributions of the side connections are incorporated in the self-energy term, which can be obtained by an iterative procedure [21, 22].

The transmission probability and the density of states (DOS) are calculated using the formalism introduced in Refs. [23, 24]. The local density of states (LDOS) refers to the distribution of electronic states at a specific energy level. In short, the equations to obtain these quantities are as follows:

The Green's function of the junction is defined as follows:

$$\boldsymbol{G}(E) = [(E + i\eta)\mathbf{I} - \boldsymbol{H} - \boldsymbol{\Sigma}_{\mathrm{L}}(E) - \boldsymbol{\Sigma}_{\mathrm{R}}(E)]^{-1}, \tag{1}$$

where $\eta$ is an arbitrary infinitesimal number, $\mathbf{I}$ is the identity matrix, and $\boldsymbol{H}$ is the real space Hamiltonian matrix of the device. $\boldsymbol{\Sigma}_{\mathrm{L}}(E)$ and $\boldsymbol{\Sigma}_{\mathrm{R}}(E)$ are the self-energy terms, representing the effective Hamiltonian due to the coupling of the central region with the left and right terminals, respectively. The self-energies are computed using the iterative procedure described in references [21, 22].

The electronic transmission probability is given by:

$$T_e(E) = \text{Tr}[\boldsymbol{\Gamma}_L(E)\boldsymbol{G}(E)\boldsymbol{\Gamma}_R(E)\boldsymbol{G}^\dagger(E)], \tag{2}$$

where $\boldsymbol{\Gamma}_{L(R)}$, the broadening function of the left (right) terminal, is defined as:

$$\boldsymbol{\Gamma}_{L(R)}(E) = i\left[\boldsymbol{\Sigma}_{L(R)} - \left(\boldsymbol{\Sigma}_{L(R)}\right)^\dagger\right]. \tag{3}$$

Additionally, the density of states and local density of states at energy (E) are described, respectively:

$$\text{DOS}(E) = -\frac{1}{\pi}\text{Im}\left(\text{Tr}(\boldsymbol{G}(E))\right), \tag{4}$$

$$\text{LDOS}(E)_i = -\frac{1}{\pi}\text{Im}(G(E)_{i,i}), \tag{5}$$

where $i$ is atom index. Indeed, the local density of states is a measure of the number of electronic states available at a specific energy level and location within a material.

**Nanoribbon and Vacancy Modeling**

Here, the modeling of nanoribbons and vacancies in phosphorene is described. $N_z$ZPNR ($N_a$APNR) is denoted as a ZPNR (an APNR), with the width of $N_z$ ($N_a$) defined as the number of zigzag lines (rows of atoms) across the ribbon. A monovacancy is a single atom-sized hole created by removing a phosphorus atom from a pristine PNR. For a vacancy-defected ribbon, the supercell is assumed to have a width of $N_z$ or $N_a$ and a length of $N_l$, which equals the number of chains, so that the periodicity is only along the ribbon length. Within a supercell, the position of a vacancy is uniquely identified by $V_{\alpha,\beta}$, where $\alpha$ and $\beta$ represent the indices of the corresponding chain and the removed atom within a chain, respectively.

The focus of this study is to investigate the effect of vacancies on the electronic properties of quantum dots in phosphorene-based junctions. The size of the scattering regions is chosen to allow the localized states to act independently, minimizing their interactions. However, to simplify the calculations and to allow the spatial representation of the electronic distributions, the sizes are kept as small as possible. Interestingly, the results of this study can be generalized to larger sizes.

**RESULTS AND DISCUSSION**

We first study the electronic structures of a ZPNR and vacancy-defected PNRs by an inspection of figures 1(b)-1(d). Figures 1(b) and 1(c), respectively, show the in-gap bands of 16ZPNR composed of zigzag-edge states and the vacancy-induced bands of 21APNR containing two monovacancies of $V_{2,14}$ and $V_{2,30}$ per supercell with $N_l = 3$. Both types of ribbons exhibit nearly degenerate bands (two-fold). Note that a monovacancy gives rise to a single localized in-gap level. Namely, there is a correspondence between the number of vacancies in a supercell and the number of levels created within the gap. The figures reveal that the vacancy-induced levels exhibit a flatter dispersion relation compared to the edge modes of zigzag-edged ribbon. This is evident in the effective masses of electrons and holes, which indicates that the effective mass of carriers near the zigzag edges is smaller than that around the monovacancies. Figure 1(d) displays the spatial LDOS of the defected 16ZPNR integrated over the in-gap states for a supercell (the periodicity is along the zigzag edges). The figure shows that the electron densities are spatially localized at the two zigzag edges as well as close to the vacancy sites, and their distributions have mirror symmetry with respect to their central axis. Moreover, the major contribution to the vacancy-induced state of a PNR comes from the neighboring atom of the vacancy site located in the opposite layers.

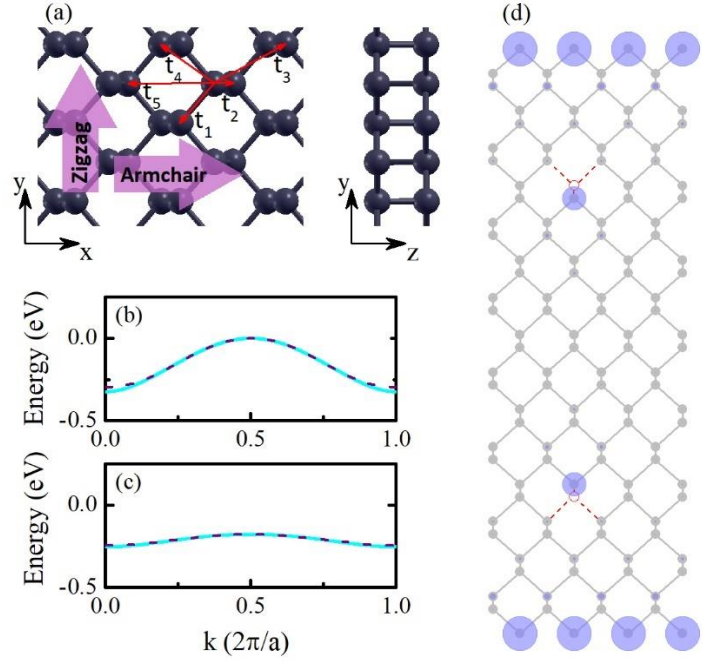

Figure 1. (a) The lattice geometry of 2D phosphorene. The x and y axes are the armchair and the zigzag directions, respectively. The different hopping parameters are indicated as well. The energy dispersion of the in-gap states of (b) 16-ZPNR, (c) defected 21-APNR containing two monovacancies of $V_{2,14}$ and $V_{2,30}$ per supercell with $N_l = 3$. (d) The LDOS of defected 16-ZPNR containing two monovacancies of $V_{2,8}$ and $V_{2,25}$ per supercell with $N_l = 4$, integrated over the in-gap states. The removed atoms and bonds are indicated with the red unfilled circles and dashed lines, respectively.

Given that the configuration with the lowest total energy might involve changes in the positions of atoms neighboring a vacancy, a systematic study was carried out on the electronic structure of nanoribbons with vacancies and various local atomic geometries. The results are shown in figures 2(a) and 2(b). $\Delta_{\alpha,\beta}$ represents the displacement of the atom adjacent to the vacancy, expressed as a percentage of the bond length, where $\alpha$ and $\beta$ represent the indices of the corresponding chain and the displaced atom, respectively. Figure 2(a) depicts the in-gap states of a defected structure where it is assumed that all three neighboring atoms of the vacancy are displaced. Figure 2(b) presents the case where only one of the neighboring atoms of the vacancy is considered to be displaced. The figures indicate that the in-gap states exhibit minimal sensitivity to changes in the geometry around the vacancy site. This result highlights the highly localized nature of these states and their relative independence from small structural variations.

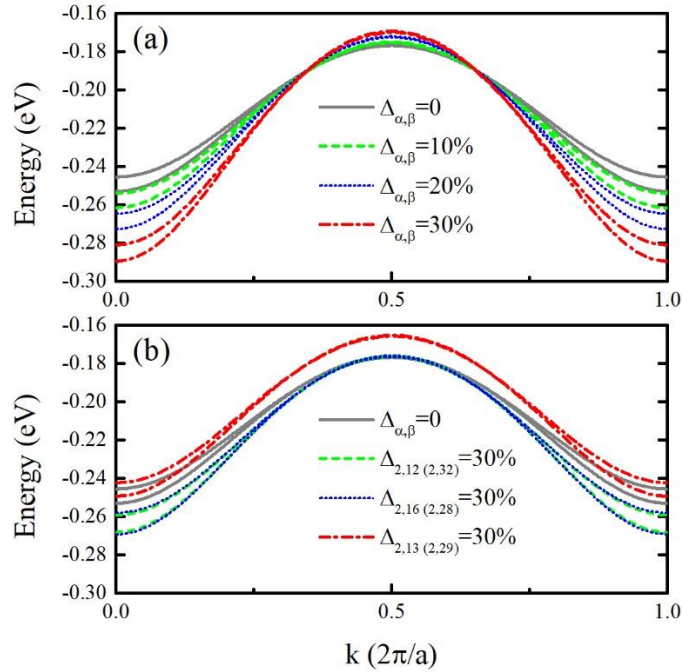

**Figure 2. The evolution of the energy dispersion of the in-gap states corresponding to figure 1(c) by various atomic displacements around the vacancy site for the case where (a) all three neighboring atoms are assumed to be displaced, (b) only one of the neighboring atoms of the vacancy is assumed to be displaced.**

We further investigate the impact of vacancies at different lattice sites on the electronic structure of the ribbon. In this study, the inter-vacancy distance is assumed to be large enough to treat vacancies as independent defects. Figure 3 illustrates the localized states induced by a vacancy of $V_{2,30}$ as a reference and another vacancy at different lattice positions within the supercell of 21APNR with $N_l = 3$. The results indicate that varying the position of the second vacancy leads to minor changes in the in-gap modes. Thus, considering vacancy independence, the precise location of vacancies has a negligible impact on the electronic properties of the system.

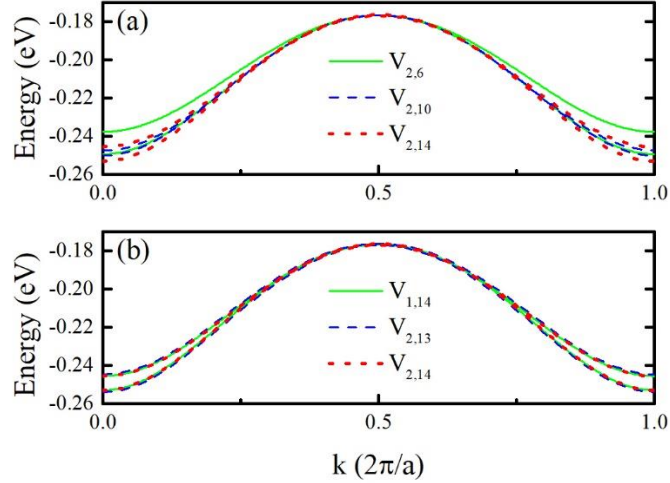

**Figure 3. The in-gap states of 21APNR induced by a vacancy of $V_{2,30}$ as a reference and another vacancy at different lattice positions within the supercell with $N_l = 3$. (a) The y-component, (b) the x-component of the inter-vacancy distance are different.**

Here, we present efficient QDs trapped in various semiconductor/metal/semiconductor junctions tailored from phosphorene. Specifically, we investigate three distinct models. As schematically shown in the inset of figure 4(a), the first model is based on a straight junction of 21APNR, including a vacancy-defected region sandwiched by perfect ribbons. The subsequent two junctions adopt a bent model. We utilize either perfect or vacancy-defected 11ZPNR and two semi-infinite 7APNR as the building blocks of the S-shaped QD devices, as depicted in the inset of figure 4(b) and the top inset of figure 4(c), respectively. Figures 4(a)-4(c) show the spin-resolved transmission spectra of the junctions subjected to the FM exchange field of $M = 0.1$ eV, corresponding to their insets. The transmission of pristine 21APNR having a step-like behavior [25] is slightly affected by the vacancies. In the cases of S-shaped systems, there are resonant tunneling peaks in the transmission spectra arising from the scattering of electrons passing through the bent [26, 27]. However, as shown in the figures, all these junctions are gapped, and the exchange field induces opposite spin polarization at the two edges of the energy gap.

In our previous work, we demonstrated that zigzag $MoS_2$ nanoribbons can exhibit half-metallic behavior in the presence of a transverse electric field and an exchange field [28]. Our calculations on PNRs reveal that in-gap states play a critical role in enabling these nanostructures to achieve half-metallicity under similar conditions.

Figures 4(d)-4(f) display the DOS spectra of the junctions corresponding to figures 4(a)-4(c), respectively. With the discrete levels within the energy range of the band gap, these junctions clearly act as QD devices and can completely confine electronic states induced by the topological structure of the system. In the case of the straight junction of 21APNR, there are two close peaks within the gap (see figures 4(a) and 4(d)) corresponding to the two flat bands of 21APNR induced by two vacancies (see figure 1(c)). In such a system, the presence of a phosphorus vacancy is crucial to the formation of the QD, and the vacancy behaves like an electron trap. A similar behavior has also been demonstrated in a graphene-based device in which electrons are trapped by vortex currents around the carbon vacancy [29].

We demonstrated that our model can be a spin-selective quantum dot using an exchange field. It is worth noting that vacancies in phosphorene can induce local spin polarization leading to the formation of localized magnetic moments, as discussed in Ref. [12]. This effect is due to the imbalance in the number of electrons around the vacancy, which can result in a net spin polarization. Although our simplified model did not explicitly account for this phenomenon, however, vacancies themselves can inherently create spin polarization, potentially enabling the realization of spin-selective quantum dots even without any exchange field.

The localized states of the perfect S-shaped junction stem from the zigzag edges of the middle region, including more peaks. A comparison between the DOS spectra of two S-shaped systems indicates that the vacancy-induced peaks are detached from the edge states and emerge beyond the pristine peaks, which are marked by black arrows in figure 4(f). We illustrate the spin-dependent energy levels of the isolated middle region of the defected S-shaped system in the bottom inset of figure 4(c). The well-separated levels, marked by arrows, are actually induced by vacancies, and they are doubly degenerate. It is evident that the degeneracy is resolved over the contact when compared to figure 4(f). By varying the width of the middle region or the number of monovacancies, the spatial confinement and the energy level distribution can be modified accordingly. As shown in figures 4(d)-4(f), the localized states can be spin-selective using an FM exchange field. However, it should be noted that the vacancy-induced states may be inherently spin-polarized (in the absence of exchange fields), as we have discussed above.

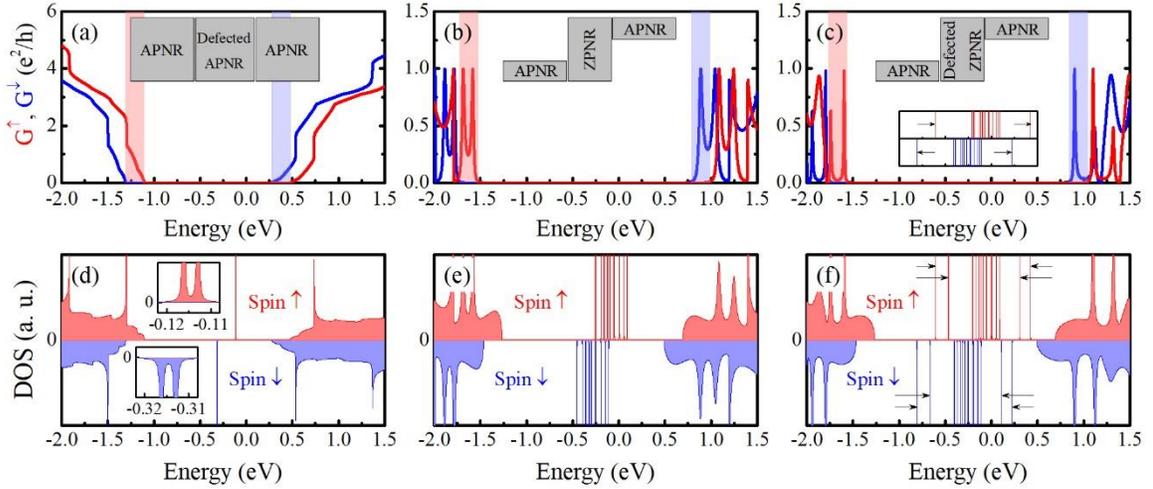

**Figure 4.** The spin-resolved transmission spectrum of (a) 21APNR with two vacancies applied to the middle region, the S-shaped junction including (b) perfect, and (c) vacancy-defected 11ZPNR connected to two semi-infinite 7APNR in the presence of FM exchange field of $M = 0.1$ eV. Schematic views of the junctions are shown in the insets of (a)-(c). In addition, the bottom inset of (c) illustrates the energy levels of the isolated middle region of the defected S-shaped system. (d)-(f) The DOS spectra of the junctions corresponding to (a)-(c). The insets of (d) are the enlarged views of the corresponding in-gap peaks.

When an in-plane electric field is applied across a PNR, significant changes in electronic structure are induced for both armchair- and zigzag-edged cases. As discussed in Ref. [30], the band gap of APNRs subjected to a transverse electric field decreases as the electric field increases and eventually closes at a critical electric field. In the case of ZPNR, the quasi-flat band is highly tunable by applying a transverse electric field, and the transmission probability is switched off at a critical electric field, and therefore, the system behaves as a field-effect transistor [31].

Now, let us apply a transverse electric field to a ribbon with armchair edges containing two monovacancies per supercell. We present the band structure of the defected 21APNR with $V_{2,14}$ and $V_{2,30}$ in the absence (the dashed lines) and in the presence (the solid lines) of the transverse electric field of $E_y = 0.01$ V/Å in figure 5(a). In the presence of the electric field, the band gap decreases, and the nearly degenerate flat bands undergo a band splitting of ~ 0.13 eV. The distribution of probability amplitude shown in figure 5(b) is associated with the degenerate in-gap mode for $k = \pi/a$ without any external field, which indicates that the electronic states are symmetrically located at the neighboring atom of each vacancy site. Figures 5(c) and 5(d) correspond to the distribution functions of the in-gap states influenced by the transverse electric field. The applied electric field breaks the symmetry and causes the electronic states associated with the two in-gap modes to localize around the opposite vacancy sites, resulting in one mode shifting upward (figure 5(c)) and the other shifting downward (figure 5(d)), while their shapes remain unchanged. When an electric field is transversely applied to a defected APNR pointing from bottom edge to top, the electrostatic potential at the upper and lower edges is thus raised and lowered, respectively. The potential difference between the two vacancy sites ($\propto lE_y$, where $l$ is the distance between the vacancy sites along the ribbon width) lifts the degeneracy of the vacancy-induced in-gap modes and pushes the bands away from the Fermi level to higher and lower energies. What we observe here is a giant Stark effect [32].

We now analyze the relationship between the spin-resolved transport properties of the 21APNR QD (displayed in the inset of figure 4(a)) and the transverse electric field in the presence of $M = 0.1$ eV. The evolution of the vacancy-induced localized states in the DOS spectrum with the field of $E_y = 0.01$ V/Å is shown in figure 5(e) by the arrows. When the electric field is applied, the overlap between the two localized states is eliminated and the peaks split perfectly by the energy of ~ 0.13 eV, in accordance with the electric field-induced band-splitting of the defected 21APNR. Moreover, it is possible to remove the spin degeneracy using an FM exchange field and to create spin-selective localized states. Figure 5(f) illustrates the energy gap and the splitting of the localized states of the 21APNR QD as functions of the transverse electric field. It shows that the energy gap of the junction gradually decreases, while the splitting of the localized states increases as the field increases. These results can be applied to the development of electronic devices with improved performance.

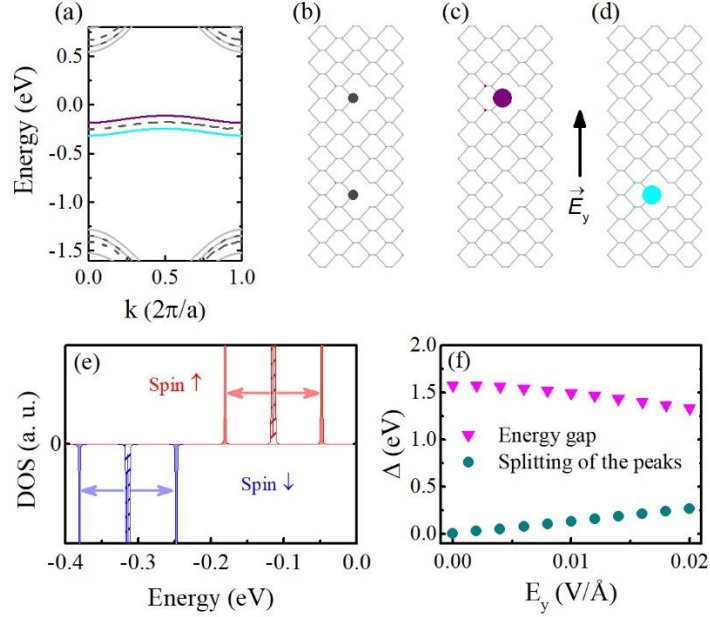

**Figure 5.** (a) The band structure of the defected 21APNR in the absence (the dashed lines) and in the presence (the solid lines) of $E_y = 0.01$ V/Å. The distribution of probability amplitude of (b) the degenerate in-gap mode, (c) and (d) the in-gap states influenced by $E_y = 0.01$ V/Å, for $k = \pi/a$. The direction of the transverse electric field is shown in (c), oriented from bottom to top. (e) The spin-resolved DOS spectra of the 21APNR-based QD subjected to $M = 0.1$ eV in the absence (with the hatched region between the curve and the horizontal axis) and in the presence of $E_y = 0.01$ V/Å (with the shaded region between the curve and the horizontal axis). (f) The energy gap and the splitting of the localized states of the 21APNR QD as functions of transverse electric field.

Figures 6(a)-6(d) display the results of in-gap states modification for the perfect S-shaped junction (illustrated in the inset of figure 4(b)) by the transverse electric field and in the presence of the FM exchange field of $M = 0.1$ eV. As the electric field increases, the band gap decreases (not shown in the figures), and the overlap between the levels tends to vanish, which makes the localized states more distinguishable. Indeed, compared to the DOS spectrum without electric field (figure 6(a)), the degenerate localized states under the electric field gradually experience a certain degree of splitting, leading a spreading of the states over a wider range of energy.

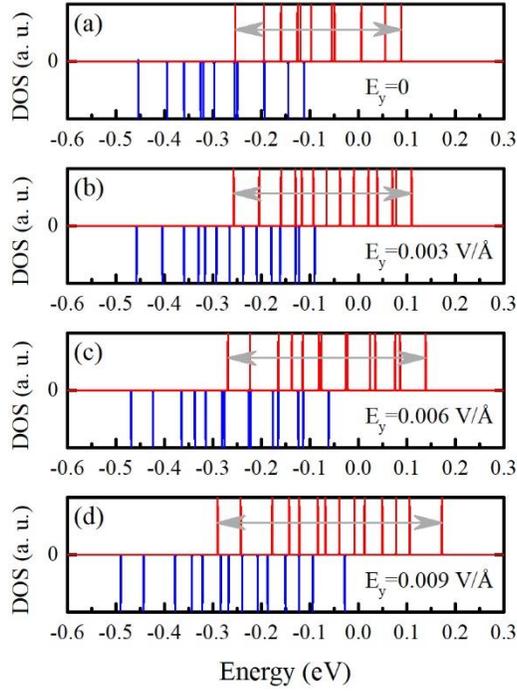

**Figure 6. (a)-(d) The evolution of the DOS spectrum of the perfect S-shaped junction by the transverse electric field and in the presence of $M = 0.1$ eV.**

**CONCLUSION**

We found that monovacancies in a PNR create localized energy levels within the band gap. Given this result, we were able to establish a design framework for a phosphorene-based QD by introducing vacancies in the middle region of an armchair-edged straight junction. In addition, we investigated QD junctions adopting a bent configuration. We demonstrated that by varying the width of the middle region or the number of monovacancies, the spatial confinement and the distribution of sharp peaks within the band gap can be modified accordingly. Moreover, transverse electric fields were found to be used to engineer the energy levels of the QDs, enabling a wide range of applications in quantum electronics. Additionally, the calculations showed that the localized states of the QD devices can be spin-selective using an FM exchange field. This finding may pave the way for the development of more powerful quantum computers.

## AUTHOR DECLARATIONS

### Conflict of Interest

The authors have no conflicts to disclose.

### Author Contributions

**Maryam Mahdavifar:** Conceptualization (equal); Data curation (equal); Formal analysis (equal); Investigation (equal); Methodology (equal); Software (lead); Validation (equal); Visualization (equal); Writing – original draft (lead). **Farhad Khoeini:** Conceptualization (lead); Data curation (equal); Formal analysis (equal); Investigation (equal); Methodology (equal); Supervision (lead); Validation (equal); Visualization (equal); Writing – review & editing (equal).

**François M. Peeters:** Conceptualization (equal); Formal analysis (equal); Data curation (equal); Writing – review & editing (equal).

## DATA AVAILABILITY

The data that support the findings of this study are available from the corresponding authors upon reasonable request.

khoeini@znu.ac.ir

## REFERENCES

1. Castellanos-Gomez A, Vicarelli L, Prada E, Island JO, Narasimha-Acharya K, Blanter SI, et al. Isolation and characterization of few-layer black phosphorus. 2D Materials. 2014;1(2):025001.
2. Lu W, Nan H, Hong J, Chen Y, Zhu C, Liang Z, et al. Plasma-assisted fabrication of monolayer phosphorene and its Raman characterization. Nano Research. 2014;7:853-9.
3. Tran V, Soklaski R, Liang Y, Yang L. Layer-controlled band gap and anisotropic excitons in few-layer black phosphorus. Physical Review B. 2014;89(23):235319.
4. Li L, Kim J, Jin C, Ye GJ, Qiu DY, Da Jornada FH, et al. Direct observation of the layer-dependent electronic structure in phosphorene. Nature Nanotechnology. 2017;12(1):21-5.
5. Xia F, Wang H, Jia Y. Rediscovering black phosphorus as an anisotropic layered material for optoelectronics and electronics. Nature Communications. 2014;5(1):4458.
6. Li L, Yu Y, Ye GJ, Ge Q, Ou X, Wu H, et al. Black phosphorus field-effect transistors. Nature Nanotechnology. 2014;9(5):372-7.
7. Koenig SP, Doganov RA, Schmidt H, Castro Neto A, Özyilmaz B. Electric field effect in ultrathin black phosphorus. Applied Physics Letters. 2014;104:103106.
8. Liu H, Neal AT, Zhu Z, Luo Z, Xu X, Tománek D, et al. Phosphorene: an unexplored 2D semiconductor with a high hole mobility. ACS Nano. 2014;8(4):4033-41.
9. Watts MC, Picco L, Russell-Pavier FS, Cullen PL, Miller TS, Bartuś SP, et al. Production of phosphorene nanoribbons. Nature. 2019;568(7751):216-20.

10. Wang H, Song Y, Huang G, Ding F, Ma L, Tian N, et al. Seeded growth of single-crystal black phosphorus nanoribbons. Nature Materials. 2024;23(4):470-8.
11. Guo H, Lu N, Dai J, Wu X, Zeng XC. Phosphorene nanoribbons, phosphorus nanotubes, and van der Waals multilayers. The Journal of Physical Chemistry C. 2014;118(25):14051-9.
12. Srivastava P, Hembram K, Mizuseki H, Lee K-R, Han SS, Kim S. Tuning the electronic and magnetic properties of phosphorene by vacancies and adatoms. The Journal of Physical Chemistry C. 2015;119(12):6530-8.
13. Nazar ND, Peeters F, Costa Filho R, Vazifehshenas T. 8-Pmmn borophene: edge states in competition with Landau levels and local vacancy states. Physical Chemistry Chemical Physics. 2024,26:16153.
14. Bafekry A, Faraji M, Fadlallah MM, Mortazavi B, Ziabari AA, Khatibani AB, Nguyen CV, Ghergherehchi M, Gogova D. Point defects in a two-dimensional $ZnSnN_2$ nanosheet: a first-principles study on the electronic and magnetic properties. The Journal of Physical Chemistry C. 2021 Jun 3;125(23):13067-75.
15. Chico L, Sancho ML, Munoz M. Carbon-nanotube-based quantum dot. Physical Review Letters. 1998;81(6):1278.
16. Wang Z, Shi Q, Li Q, Wang X, Hou J, Zheng H, et al. Z-shaped graphene nanoribbon quantum dot device. Applied Physics Letters. 2007;91(5):053109.
17. Recher P, Sukhorukov EV, Loss D. Quantum dot as spin filter and spin memory. Physical Review Letters. 2000;85(9):1962.
18. Thakur T, Peeters FM, Szafran B. Electrical manipulation of the spins in phosphorene double quantum dots. Scientific Reports. 2024 Aug 16;14(1):18966.
19. Rudenko AN, Katsnelson MI. Quasiparticle band structure and tight-binding model for single-and bilayer black phosphorus. Physical Review B. 2014;89(20):201408.
20. Taghizadeh Sisakht E, Fazileh F, Zare M, Zarenia M, Peeters F. Strain-induced topological phase transition in phosphorene and in phosphorene nanoribbons. Physical Review B. 2016;94(8):085417.
21. Sancho ML, Sancho JL, Rubio J. Quick iterative scheme for the calculation of transfer matrices: application to Mo (100). Journal of Physics F: Metal Physics. 1984;14(5):1205.
22. Sancho ML, Sancho JL, Sancho JL, Rubio J. Highly convergent schemes for the calculation of bulk and surface Green functions. Journal of Physics F: Metal Physics. 1985;15(4):851.
23. Datta S. Quantum transport: atom to transistor: Cambridge university press; 2005.
24. Mahdavifar M, Khoeini F, Peeters FM. Band engineering of phosphorene/graphene van der Waals nanoribbons toward high-efficiency thermoelectric devices. Physical Review B. 2024;109(8):085409.
25. Mahdavifar M, Shekarforoush S, Khoeini F. Tunable electronic properties and electric-field-induced phase transition in phosphorene/graphene heterostructures. Journal of Physics D: Applied Physics. 2020;54(9):095108.
26. Zhang Z, Wu Z, Chang K, Peeters F. Resonant tunneling through S-and U-shaped graphene nanoribbons. Nanotechnology. 2009;20(41):415203.
27. Mahdavifar M, Khoeini F. Highly tunable charge and spin transport in silicene junctions: phase transitions and half-metallic states. Nanotechnology. 2018;29(32):325203.
28. Khoeini F, Shakouri K, Peeters F. Peculiar half-metallic state in zigzag nanoribbons of $MoS_2$: Spin filtering. Physical Review B. 2016;94(12):125412.
29. Shen L, Zeng M, Li S, Sullivan MB, Feng YP. Electron transmission modes in electrically biased graphene nanoribbons and their effects on device performance. Physical Review B—Condensed Matter and Materials Physics. 2012;86(11):115419.
30. Sisakht ET, Zare MH, Fazileh F. Scaling laws of band gaps of phosphorene nanoribbons: A tight-binding calculation. Physical Review B. 2015;91(8):085409.
31. Ezawa M. Topological origin of quasi-flat edge band in phosphorene. New Journal of Physics. 2014;16(11):115004.
32. Zheng F, Liu Z, Wu J, Duan W, Gu B-L. Scaling law of the giant Stark effect in boron nitride nanoribbons and nanotubes. Physical Review B—Condensed Matter and Materials Physics. 2008;78(8):085423.